# Electromagnetic Side-Channel Vulnerability in QKD Equipment

**Mikio Fujiwara[1], Katsumi Fujii[1], Fumie Ono[1], Masakazu Ono[1,2], and Akihisa Tomita[1]**

[1]National Institute of Information and Communications Technology, Nukui-kita, Koganei, Tokyo 184-8795 Japan

[2] NEC Corporation, 1753 Shimonumabe, Nakahara-ku, Kawasaki 211-8666, Japan

Corresponding author: Mikio Fujiwara (e-mail: fujwara@nict.go.jp).

This paragraph of the first footnote will contain support information, including sponsor and financial support acknowledgment. For example, "This work was supported in part by the U.S. Department of Commerce under Grant BS123456."

**ABSTRACT** Although the unconditional security of quantum key distribution (QKD) protocol has been mathematically proven, actual QKD network nodes may be vulnerable to side-channel attacks such as ElectroMagnetic (EM) wave leakage. In this study, we measured EM emissions from a prototype QKD device implementing a time-bin decoy BB84 protocol with Y- and Z- basis. We measured the EM spectra and waveforms at distances of 0 m and 3 m from the transmitter and receiver inside an anechoic chamber. The results showed that the transmitter was the primary source of EM wave leakage, and distinct differences were observed between the quantum states generated (Y0, Y1, Z0, Z1) at the third harmonic of the 1241.6 MHz clock. Furthermore, larger EM signals were observed for the Z basis than for the Y basis. These state-dependent EM signals suggest a vulnerability to side-channel attacks that extract qubit information from the EM wave leakage. On the other hand, closing the EM-shielding rack resulted in an average attenuation of 21.2dB (ranging from 12.7 to 36.6dB), demonstrating that EM wave leakage other than clock-frequency components can be effectively suppressed to below −100 dBm. The findings of this study indicate that side-channel risks exist in QKD devices, but they can be reduced through appropriate EM shielding and device design. This study provides important guidelines for establishing security standards for trusted nodes in practical QKD networks.

**INDEX TERMS** ElectroMagnetic side-channel attack, Quantum Key Distribution (QKD), trusted node, electromagnetic shielding, BB84 protocol, decoy method, implementation security

## I. INTRODUCTION

Cryptographic technology has become indispensable for ensuring the confidentiality of information in modern society. At the same time, cryptanalysis techniques continue to advance; consequently, cryptographic algorithms have historically undergone generational transitions roughly every 20–30 years as they become compromised [1]. This trend highlights the risk of using computationally secure cryptography to protect information, such as genome data, that requires century-level confidentiality from the viewpoint of privacy protection.

Quantum Key Distribution (QKD) allows two remote users to share keys that are protected by the laws of physics, not by the computational complexity [2,3]. Therefore, information encrypted using a key shared by QKD with an information-theoretically secure protocol (e.g., the Vernam one-time pad) cannot be compromised by any advances in mathematical method or computational technology. This property has driven the development of QKD devices, and QKD Networks (QKDNs) with key-management functionality. Field demonstrations of QKDN have been conducted in various locations around the world [4-8]. Furthermore, quantum secure cloud systems, which combine QKDNs with secret sharing, quantum computers, and high-performance computers to realize highly confidential information transmission, storage, and processing, have also been proposed and demonstrated [9,10].

Although QKD protocols have been mathematically proven to be secure against any eavesdropping attack, providing a security guarantee for an actual QKD device is a nontrivial task [11,12]. There has been extensive research on eavesdropping methods and countermeasures targeting QKD devices, and there are ongoing international standardization efforts [13-18]. Nevertheless, to guarantee the security of a QKDN, it is essential not only to ensure the security of the device but also to secure the network nodes. In other words, although QKD guarantees the unconditional security of the communication channel, operational deficiencies at the node where the device is installed may compromise key information. Therefore, establishing node operational policies is particularly important to realize trusted nodes. These policies include security assessments, clarification of demarcation points, and development of implementation guidelines.

Since key information is classical once it resides at a node, secure implementations are required to protect the key even if an adversary with a certain ability tries to intrude. Implementation requires physical isolation, cybersecurity measures, and countermeasures against side-channel attacks similar to those used to protect ICT devices [19–22]. However, it should be noted that because the origins of side-channel leakage in QKD devices may differ from those in conventional devices, evaluation using actual QKD equipment is necessary [23]. In this paper, we focus on ElectroMagnetic (EM) emissions from QKD devices, which is generally difficult to evaluate yet is regarded as critically important in security-sensitive applications [24-26]. In particular, QKD transmitters may radiate EM emissions used for state generation, it results in leaking key information. Because such measurement is passive in nature, this attack cannot be detected, even if the QKD devices are equipped with countermeasures against side-channel attacks on the quantum channel [17]. Thus, the EM emission from the QKD transmitter will be a serious threat.

Experimental investigations of EM wave leakage from QKD devices have been reported previously [27,28]. However, the reported setting is not realistic for devices protected by physical security measures or operating under public scrutiny, because those studies placed EM probes inside the device and observed EM emissions from the circuit board directly. Probe-based attacks inside the device can be detected mechanically or electrically through device-level countermeasures, moreover, these attacks would be extremely difficult to carry out under public scrutiny. On the other hand, the measurement attempts from the outside are hardly detected. Thus, the latter makes a more realistic and severe threat. For a realistic attacker, it is necessary to observe leaking EM emissions from outside the device without disassembling it. There are few reports of EM measurement using an external antenna, which represents One report targeted a receiver mimicking a QKD device [29]. However, no studies have reported eavesdropping on a real QKD transmitter from outside of its housing.

In this report, we measured the EM emissions of a prototype QKD device with no countermeasures for EM wave leakage. We observed state-dependent EM spectra and time waveforms. These results suggest the potential for a side-channel attack to exploit EM wave leakage. We also demonstrated that placing the device in a shielded rack reduces EM wave leakage, which makes it difficult to extract key information. This result demonstrates that appropriate measures, such as EM wave shielding, can mitigate the risk of side-channel attacks. This is an essential condition for establishing implementation guidelines for Trusted Nodes in QKDNs.

## II. EXPERIMENT

The QKD device used in this experiment is based on the same design as the QKD prototype developed by NEC for the Tokyo QKD Network [5], which implements the time-bin decoy Bennett–Brassard 1984 (BB84) protocol [30]. Because this prototype is primarily intended to verify the operations of functional blocks. Since its enclosure has not been intentionally designed to prevent EM wave leakage, it is suited to examine the side-channel attacks caused by such leakage.

The configuration of the quantum communication section of the QKD transmitter is shown in Fig. 1(A). A gain-switched laser generates light a pulse at a repetition frequency of 1241.6 MHz with an 80-ps pulse duration. The pulse intensity is modulated by a Dual-electrode Mach–Zehnder lithium niobate Modulator (DMZM) to Signal (S), Decoy (D), and Vacuum (V) levels for the implementation of the decoy method [31-33]. The resulting pulse is then converted into a double-pulse of equal amplitude separated by 400 ps using an Asymmetric Mach–Zehnder Interferometer (AMZI) Planar Lightwave Circuit (PLC) [34]. By applying intensity and phase modulation to the double-pulse using a second DMZM, the one of the four states {Y0, Y1, Z0, Z1} used in the BB84 protocol is generated, where Y and Z denote the basis and 0 and 1 denote the bit value. For example, Y0 ϕ_1the state with bit value 0 of the Y basis. Information on the pulse intensity (S, D, V) and on the qubit states {Y0, Y1, Z0, Z1} are separately encoded onto the optical pulse states by binary random number signals generated by physical random number generators. The signals are amplified by modulation drivers and applied to DMZMs. EM waves may be radiated in the electric signal path from the random number generators to the optical modulators.

In this experiment, we focus on qubit information. States are generated using the relation between the input optical field $E_{in}$ and output field $E_{out}$ of the DMZM described by

$$E_{out} = e^{i(\phi_1+\phi_2)/2} \cos\frac{\phi_1+\phi_2}{2} E_{in}. \quad (1)$$

As illustrated in Fig. 1(B), by setting $\phi_1$= 0° or 90° in a bit period and applying phases $\phi_2$= ( 0°, 180°) or (180°, 0°) to the components of a double-pulse, the four states can be generated using two binary signals. This driving sequence is designed to yield the symmetric occurrence of marks and spaces, which suppresses the drift of the modulation characteristics in lithium niobate (LN) modulators.

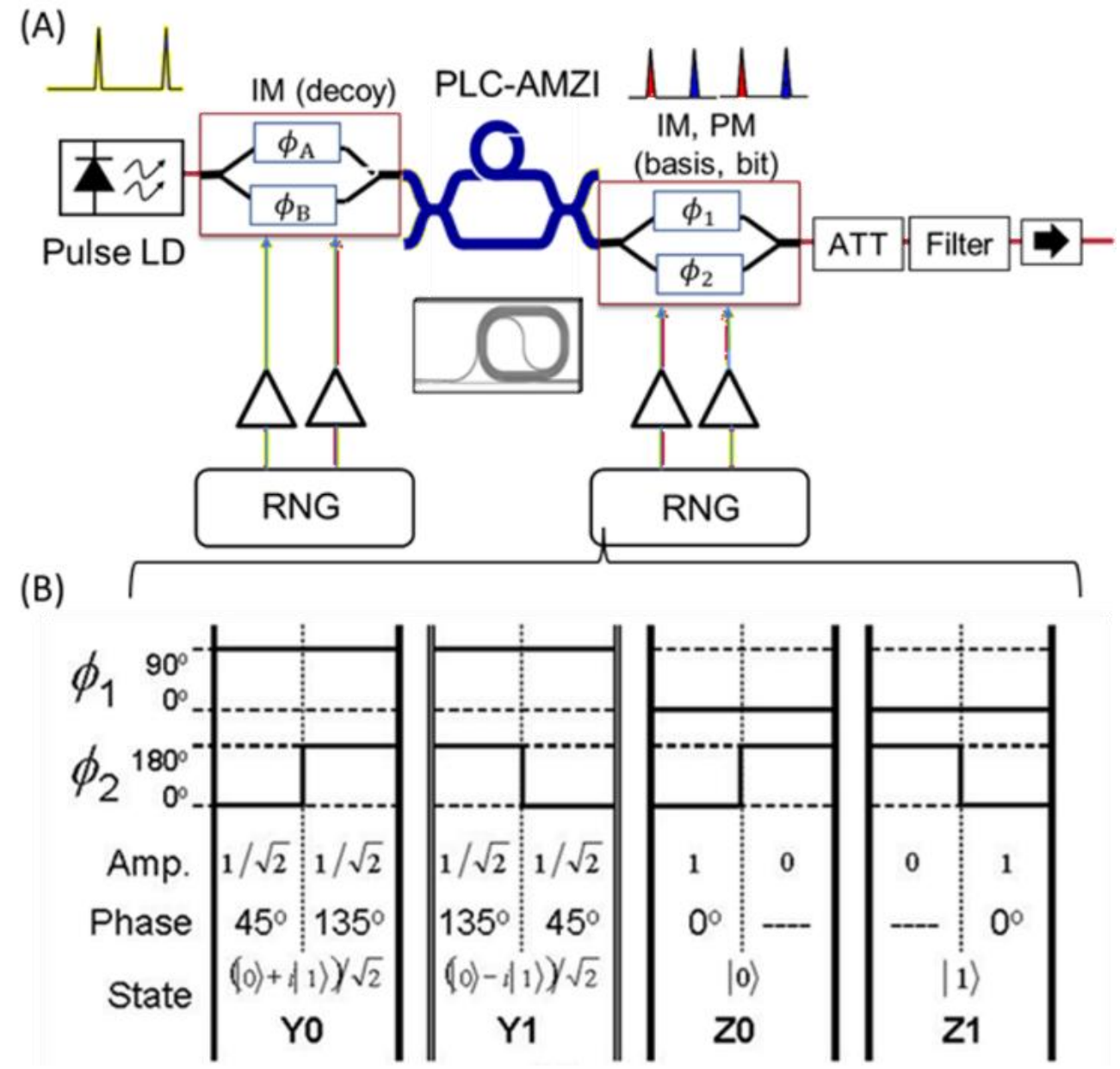

**FIGURE 1. The QKD transmitter prototype used in the experiment. (A) Schematic diagram of quantum communication section. Two lithium niobate modulators are driven by control pulses. (B) Pulse sequence for the state generation. RNG is a random number generator.**

On the other hand, the QKD receiver uses passive basis selection and requires no electrical modulation. Therefore, most components in the quantum communication section of the receiver will not generate EM radiation related to key information. Only avalanche photodiodes (APDs) used for photon detection in Gated Geiger mode may emit EM wave when the avalanche breakdown occurs. Because the high-frequency characteristics of APDs vary among individual devices, different signals may be generated depending on which APD undergoes breakdown [35]. EM waves may be generated in electronic circuits after the photon detection, raising the possibility that received signals could be estimated from EM emissions.

In this experiment, we measured the EM emission from the QKD transmitter and receiver placed them in EM-shielded racks. We observed the shielding effect of the racks by comparing the EM emissions when the rack-doors were opened and closed. As shown in Fig. 2, the racks containing the QKD devices and the antennas were installed inside an anechoic chamber to prevent the influence of external EM sources. EM emissions from the QKD device were measured in three frequency bands: 10–300 MHz, 300–2000 MHz, and 1000–6000 MHz. The antennas used for each band were a biconical antenna (Schwarzbeck, BBA9106), a log-periodic antenna (Schwarzbeck, UHALP9108A), and a horn antenna (Schwarzbeck, BBHA9120B), respectively. An omnidirectional antenna with a bandwidth of 300 MHz–8 GHz (Aaronia AG, OmniLOG30800) was also used for measurements in the 1000–6000 MHz range. The received EM signals were observed as frequency spectra and time waveforms using a signal analyzer (Keysight, N9040B) and a 10-GHz oscilloscope (Keysight, UXR104B), respectively. The resolution bandwidth (RBW) and video bandwidth (VBW) of the spectrum analyzer were set to 1 kHz and 10 kHz, respectively. EM measurements were performed with the antenna placed at 3 m from the rack; for comparison, measurements were also performed at zero distance.

Measurements were first performed with the door of the EM shielding rack open to confirm EM wave leakage, and then with the rack door closed to verify the effects of EM shielding. During the experiment, the optical state was either fixed to one of the four states {Y0, Y1, Z0, Z1}, or set to a repetitive pattern consisting of a 32-bit fixed-state header followed by 96 bits of random states. The latter pattern simulates the transmitted light during actual operation.

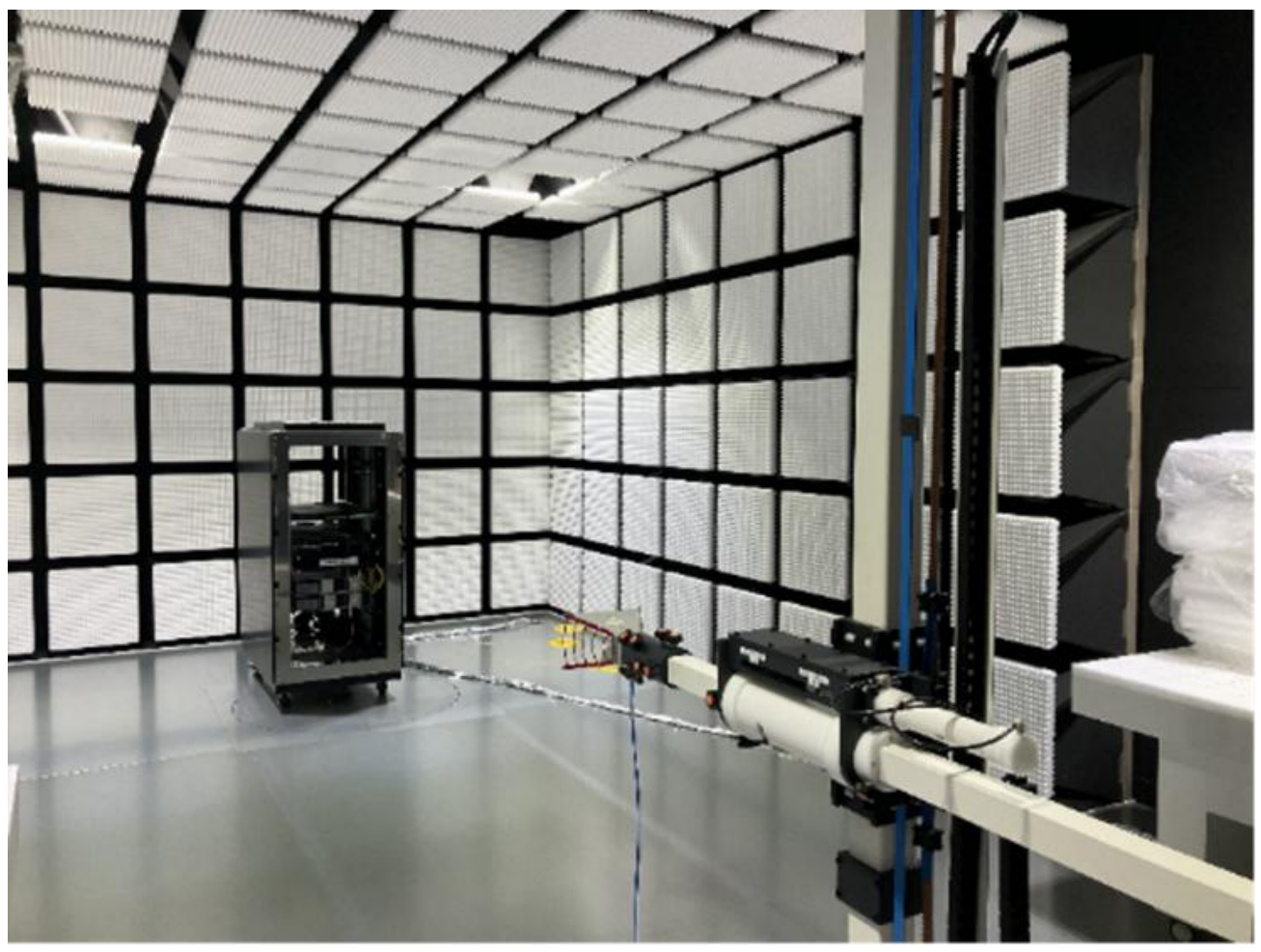

**FIGURE 2. The QKD module (transmitter) placed in a shield room. Antennas were placed at 3 m from the QKD module.**

## III. RESULTS AND DISCUSSION

### *A. CONTRIBUTION OF TRANSMITTER AND RECEIVER*

We first compare EM wave leakages from the transmitter and receiver. Figure 3 shows the observed EM spectra using the repetitive pattern when the rack doors were opened. The vertical polarization component of EM in the 1000–6000 MHz band was measured using the horn antenna. The spectra are smoothed by a 10-point moving average. Figure 3(A) shows the EM spectrum where both transmitter and receiver were powered on; Fig. 3(B) shows the case where the receiver was powered off; and Fig. 3(C) shows the case where the receiver was powered off, and the transmitter was in an idle state (modulation for state stopped). In Fig. 3 (C), almost no EM radiation was observed. Comparing Fig. 3 (A) and Fig. 3 (B), though the contributions from the receiver were observed at the clock frequency and its second harmonic, it is evident that EM wave leakage originating from the transmitter dominated. This is because the receiver does not require large-amplitude high-frequency signal modulation in its circuit configuration, resulting in lower EM wave generation intensity. Accordingly, all subsequent measurements focus on the transmitter.

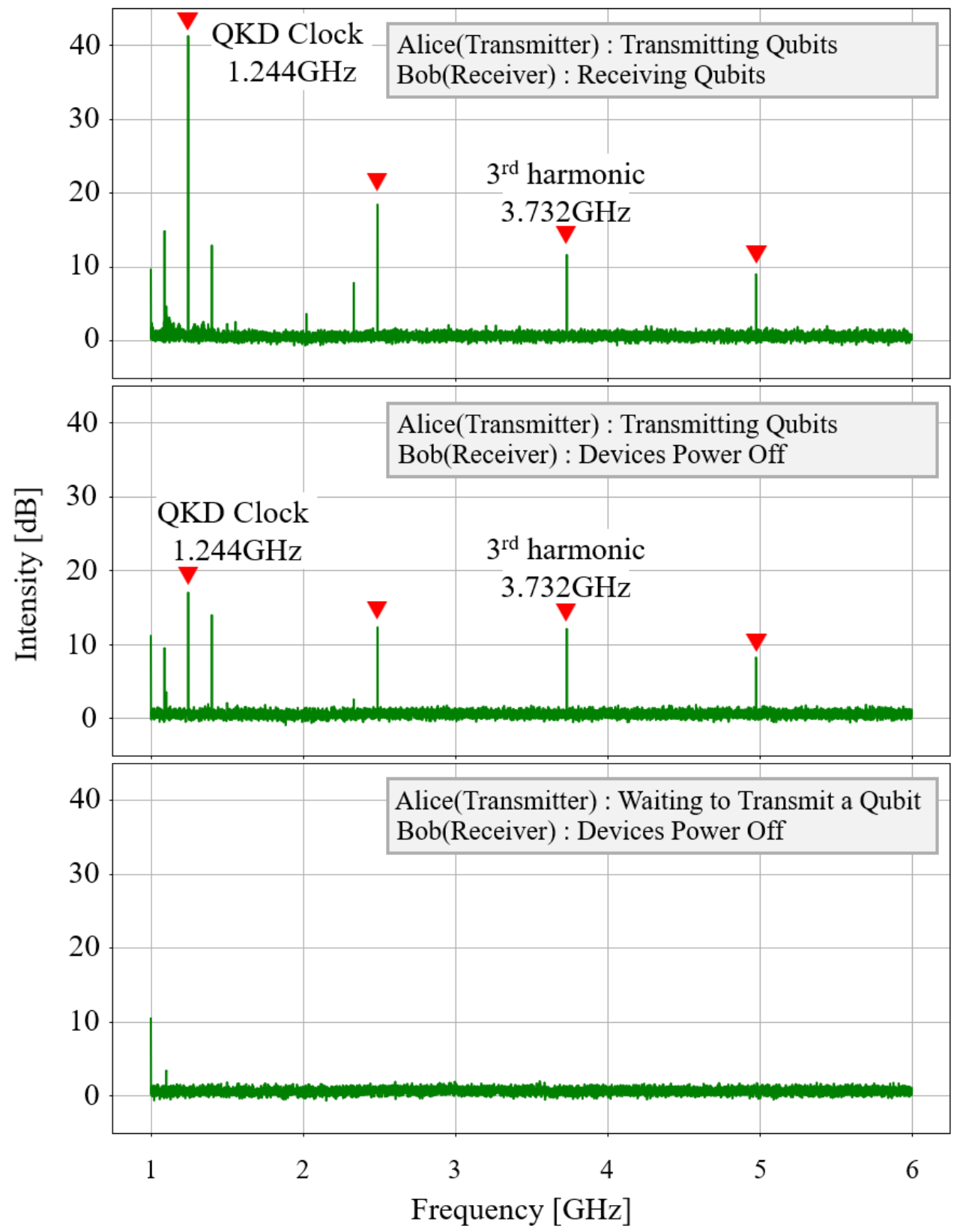

**FIGURE 3. Electromagnetic wave spectra from the QKD transmitter (Alice) and the receiver (Bob). Triangles indicate clock (1241.6 MHz) and its harmonics. The traces were obtained through the utilization of a ten-point moving average. The intensity spectra were measured with reference to the spectrum when both Alice and Bob were powered OFF. Top: Both Alice and Bob were ON. Middle: Alice was ON, but Bob was OFF. Bottom: Power of Alice was ON but state modulation was stopped; Bob was OFF.**

### *B. STATE-DEPENDENT EM SPECTRA*

Figure 4 shows the measured EM spectra for fixed states at 3 m. The difference between the spectrum of each state and the Z0 (reference) spectrum is shown in blue. In the radio-frequency band of 10–300 MHz, the observed difference between the Z0 and Z1 spectra is small. A difference was observed between the Y0 and Z0 spectra in the 80–150 MHz range. The spectrum difference was similar to that can be obtained by a difference between frequency-shifted noise spectra. However, since the master clock position is unchanged, this is not attributable to a frequency-origin offset during measurement. Between Y1 and Z0 spectra, difference was observed near 20 MHz, and stronger EM radiation in the Y1 state is found near 140 MHz.

In the microwave region above 300 MHz, overall spectral differences were small, but differences appeared in specific line spectra. EM radiation near 850 MHz was detected when the transmission state was Z0 or Z1, but not in the Y0 or Y1 states. Near the clock frequency of 1241.6 MHz, stronger EM radiation was observed in the Z0 state than in the other states. Additionally, EM radiation was stronger in the Z1 state than in the Y0 and Y1 states. This finding suggests that EM radiation intensity differs between the Z and Y bases at the clock frequency. Similar observations were obtained with both the log-periodic and horn antenna, confirming that this EM radiation originates from the QKD device. Additionally, at the third harmonic of the clock frequency, stronger EM radiation was observed for states with a bit value of 0 (Y0 and Z0) than for states with a bit value of 1 (Y1 and Z1). Table I summarizes the intensity differences observed during quantum signal generation at the fundamental wave, the second harmonic, and the third harmonic.

The results presented in Figs. 3 and 4 demonstrate that EM wave leakage from QKD devices can be readily measured using commercially available equipment. Moreover, relative spectral differences were clearly observed with a sufficiently high signal-to-noise (S/N) ratio, even after accounting for the background noise of the measurement system. These findings strongly suggest that information regarding the state generation process may be inferred from the leaked EM wave emissions.

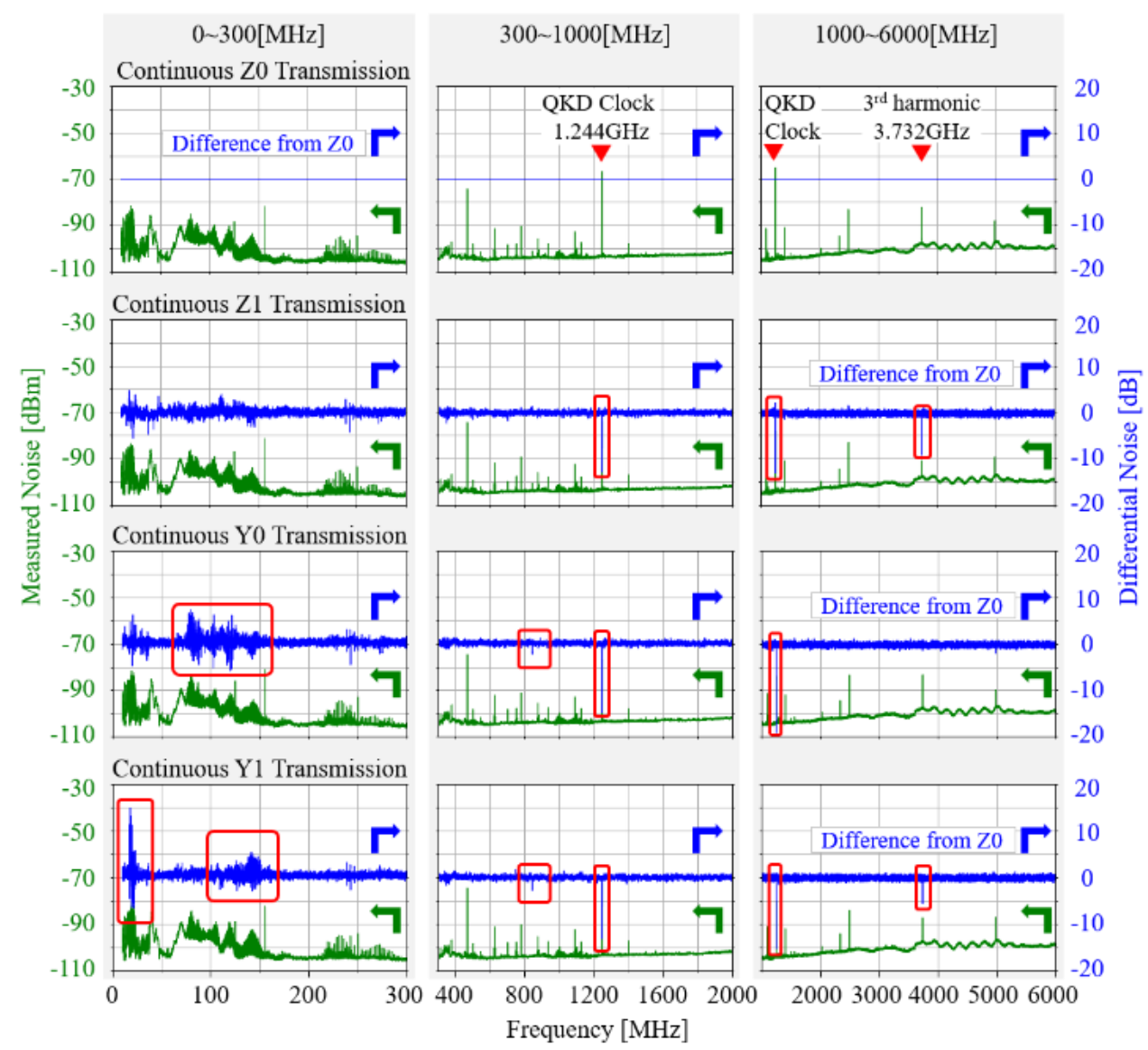

**FIGURE 4. Measured EM spectrum from the QKD transmitter (Alice) at 3 m. The frequency ranges, from left to right, are 10–300 MHz, 300–2000 MHz, and 1000–6000 MHz. The quantum states set in the transmitter are, from top to bottom, Z0, Z1, Y0, and Y1. Green represents the observed spectrum, while blue indicates the difference from the spectrum in the Z0 state. Areas showing significant differences are circled in red. The master clock (m-clock) frequency is indicated by a blue triangle, and the optical signal clock by a red triangle. No correction was made for background noise or antenna gain.**

TABLE I
INTENSITY DIFFERENCE DURING QUANTUM STATES GENERATION AT THE FUNDAMENTAL WAVE, SECOND HARMONIC, AND THIRD HARMONIC OVERVIEW OF QKD SYSTEM SPECIFICATIONS.

| **Freq. (GHz)** | **Z0** (dB) | **Z1** (dB) | **Z0-Z1** (dB) | **Y0** (dB) | **Y1** (dB) | **Y0-Y1** (dB) |
|---|---|---|---|---|---|---|
| **1.2442** | 31.8 | 19.0 | +12.8 | 13.5 | 17.0 | -3.4 |
| **2.4883** | 11.4 | 11. 6 | -0.2 | 11.1 | 10.6 | +0.5 |
| **3.7325** | 9.3 | 3.1 | +6.3 | 8.2 | 5.1 | +3.2 |

### C. TIME-DOMAIN WAVEFORM

Figure 5 shows the observed time-domain waveforms measured using a broadband omnidirectional antenna with averages taken over 16 samples at 0 m. The amplitudes of the EM signals in the Z basis were larger than those in the Y basis. Moreover, the waveforms in the Y0 and Z0 states were distorted more from sinusoidal waveforms than those in the Y1 and Z1 states. These observations are consistent with the differences observed near the clock frequency and near its third harmonic in the EM spectra. Slow amplitude modulations with periods longer than the clock period were also observed, which are considered to reflect the differences in the EM waves spectra at frequencies below the clock frequency.

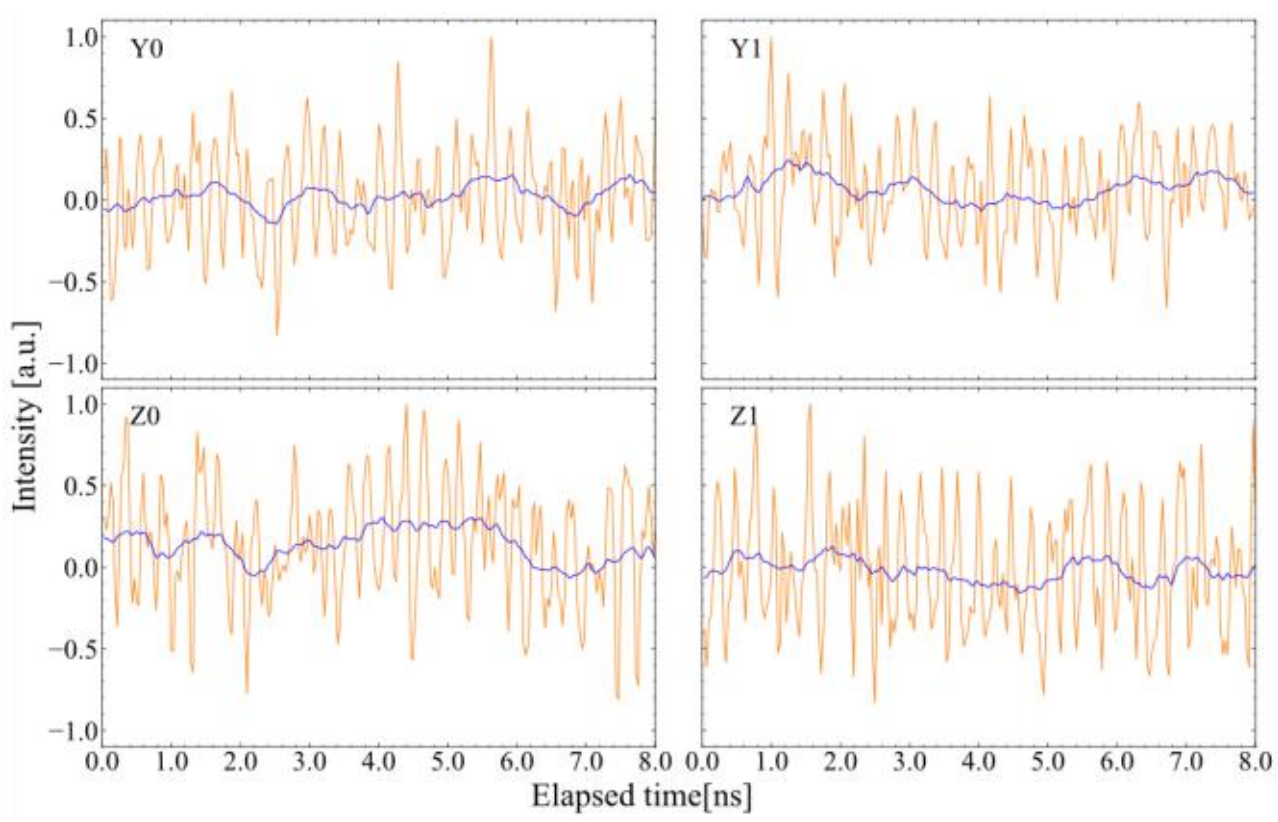

**FIGURE 5. Observed waveforms of EM field for Y0, Y1, Z0, and Z1 states at 0 m. Orange lines: real time oscilloscope traces. Blue lines: moving averages over 25 samples (corresponding to 800 ps).**

### D. SECURITY IMPLICATIONS

Based on the measurement results above, we discussed the possibility of eavesdropping exploiting EM wave leakages. In the 10–300 MHz band, clear differences between states were observed; however, because these differences imply the slower changes than the clock in time domain, it is difficult to exploit them to identify the state of individual pulses sent in the quantum communication, especially when the signal is selected randomly. Furthermore, since high-sensitivity EM interception at these frequencies requires an antenna with a large footprint on the order of wavelengths (several meters), eavesdroppers will be captured easily in a realistic setting. Nevertheless, signals in this frequency band could be used to discriminate device operating states. For example, if it is possible to distinguish the fixed-pattern state used for calibration from the random-pattern state used for key generation, this information could be exploited to improve eavesdropping. For measurements aimed at such discrimination, a smaller antenna can be used by extending the measurement time to improve the signal-to-noise ratio (SNR). Although the direct threat is limited, sufficient care must be taken to minimize the information provided to an eavesdropper.

In the frequency region above the clock frequency, a direct observation of a single EM pulse may provide a non-negligible amount of information on the states. The measurement results suggest that basis information and bit-value information can be extracted from the intensity at the clock frequency and the intensity at the third harmonic, respectively, using sophisticated methods such as Artificial Intelligence (AI)-based signal processing. Information obtained from EM leakage will impact on security guarantees.

To evaluate the accuracy of transmitter state identification based on electromagnetic leakage, a preliminary analysis was conducted using continuous wave transmissions corresponding to specific transmitter states. Features were extracted from 512 waveform blocks, each with a duration of 800 ps, and state identification models were trained using Principal Component Analysis (PCA), Linear Discriminant Analysis (LDA) with the Python packages scikit-learn, SciPy, and NumPy. The trained models were then used to classify the transmitter states in a subsequent sequence of 512 waveform blocks. As a result, state estimation accuracies of 72,7%, 76.7%, 75.8%, and 52.2% were obtained for the transmission of four states {Y0, Y1, Z0, Z1}, respectively. Although the experimental setup, which assumes continuous transmission of a single state, differs from practical operating conditions, the results indicate that transmitter states can be identified with a relatively high probability based on EM wave leakage. Furthermore, the potential risk of eavesdropping could be significantly increased under more sophisticated attack scenarios, for example, if an adversary selectively induces or exploits transmission conditions under which state estimation accuracy is particularly high.

The mechanism behind the observed differences in EM spectra and time-domain waveforms is not yet clear. However, it is considered that digital-control circuits generate state-dependent EM radiation. In contrast, the contribution of modulation drivers is thought to be limited, because a highly symmetric driving signal scheme is employed. The modulation intensity spectra of the four states are identical by design, as the signal applied to the modulator is constant for $\phi_1$, and the same rectangular wave with a half-period phase shift between bit values 0 and 1 is applied for $\phi_2$, as depicted in Fig. 1(B). This claim is consistent with the observation that differences between states in the spectra were relatively small except at the clock and its harmonics in the high-frequency region above the clock frequency.

### E. EFFECT OF EM WAVE SHIELDING

To investigate the effect of the shielded rack, EM spectra were measured at distances of 0 m and 3 m. A broadband omnidirectional antenna was used to measure the vertical polarization component from 300 MHz to 6 GHz. The optical state was set to a repetitive pattern of a 32-bit header followed by 96 random bits.

Figure 6(A) shows the spectrum measured in the immediate vicinity (0 m) of the rack with the door open. In addition to harmonics of the 1241.6 MHz clock, numerous peaks were observed below 2 GHz. Since the clock frequency is derived from a 155.77 MHz-master clock, harmonics of this frequency are considered one source of these peaks. It is also possible that fractional-frequency components of the clock are generated by the nonlinearity of the circuits and detected. The generation of sub-clock frequency EM waves may depend on the signal patterns, although the detailed identification has not been completed. Note that these spectra differ from those observed during fixed-pattern generation. This difference could be exploited to discriminate between operating modes. For example, a fixed pattern is transmitted for calibration, and a random pattern is transmitted for key generation. An eavesdropper could use the fixed pattern to learn and the random pattern to infer key information.

As shown in Fig. 6(B), moving the antenna 3 m away and keeping the rack door open resulted in a significant reduction in the intensity of the peaks below 2 GHz and their harmonics. Nevertheless, EM radiation remained at a detectable level. By closing the rack door, peaks other than the clock frequency disappeared, as shown in Fig. 6(C).

The signal intensities, other than the peak at the clock frequency, fell below −100 dBm. Figure 6(D) shows the attenuation at 0 m resulting from closing the door from the open position. When peaks in Fig. 6 (A), with intensities above -90 dBm, were compared with the peaks in Fig. 6 (C), the EM intensity was found to be attenuated by 12.7dB to 36.6dB, with an average attenuation of 21.2dB. In summary, clear EM radiation from the QKD transmitter was observed even at 3 m, but it was attenuated by approximately 20dB when the shielding rack door was closed. In conclusion, Figure 6(c) implies that the shield door enables to shut out the fundamental wave and harmonics leakage used for quantum state estimation. And this result suggests that this shield is highly effective at preventing backdoors caused by EM wave leakage.

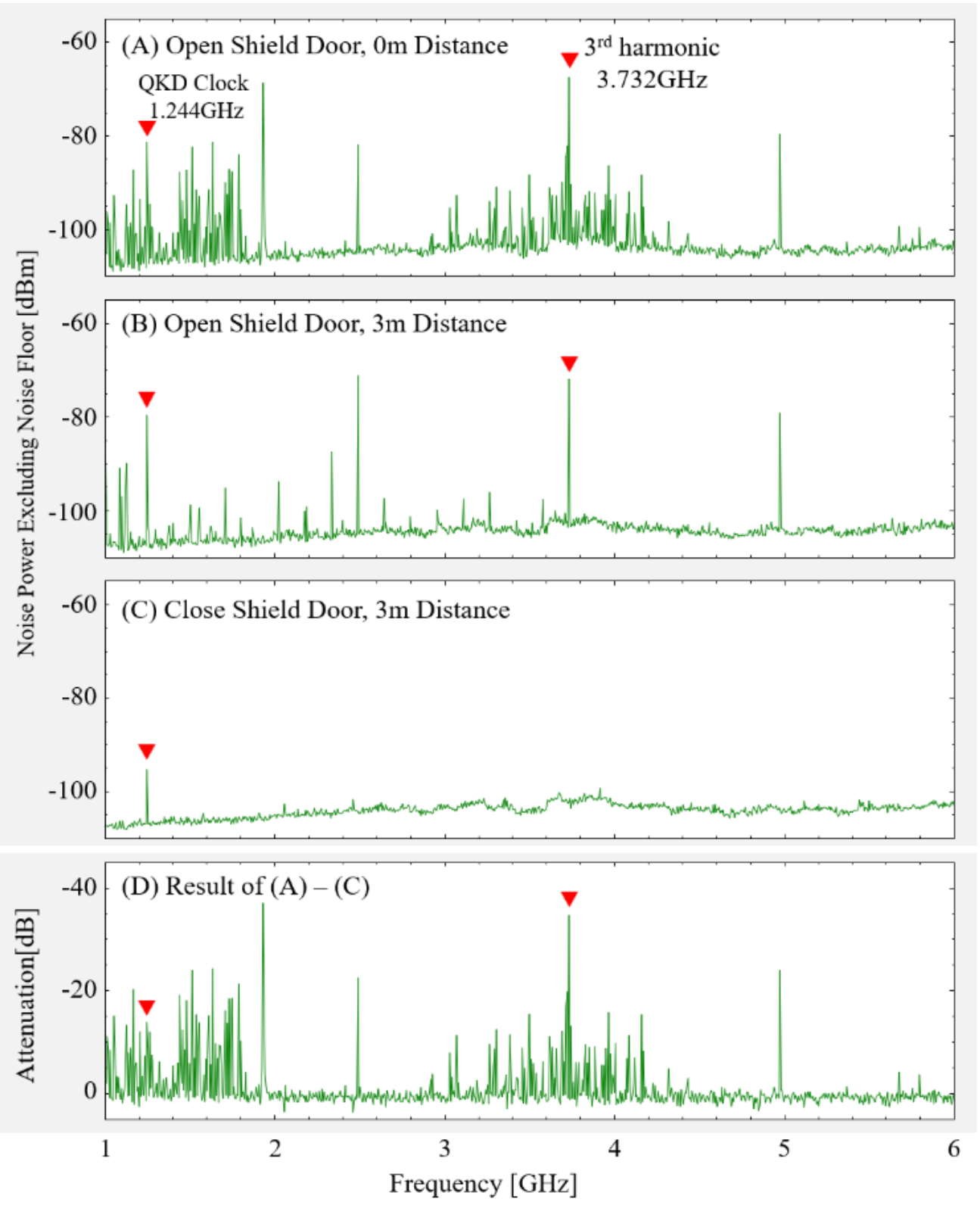

**FIGURE 6. Electromagnetic wave spectra from a QKD transmitter (Alice). (A) Spectrum measured at 0 m from the rack with door open. (B) Spectrum measured at 3 m from the rack with door open. (C) Spectrum measured at 3 m from the rack with door closed. (D) Attenuation as the difference between (A) and (C).**

## IV. CONCLUSION

In QKDN, ensuring node security requires preventing information leakage caused by external observation of the QKD devices installed in the trusted node. In this report, by using the prototype of the QKD system (not commercial one), we evaluated EM radiation from the devices. Because EM wave measurements are passive, eavesdropping detection is difficult. Therefore, eavesdropping through the EM wave leakage should be considered a realistic security threat. To ensure the security of trusted nodes, it is crucial that no information, even in small amounts, be available outside the QKD device that could lead to key information leakage through EM wave measurements. We observed non-negligible levels of EM radiation from a prototype device that had not been equipped with EM wave leakage countermeasures, when measuring the EM wave leakage at 3 m from the QKD device.

It was demonstrated that the transmitter contributes particularly to EM leakage. These results suggest that it is possible to estimate the state of the transmitted light using the leaked EM radiation. The impact is particularly severe in cases where, unlike this device, the modulation signal has low symmetry, and the transmission state can be easily estimated from the drive signal.

Countermeasures against EM wave side-channel attacks require the following: (1) suppression of EM generation within the device, (2) prevention of generated EM leakage outside the device, and (3) prevention of leaked EM from providing useful information. One possible approach to suppressing EM wave generation within the device is to use a modulator with a smaller drive voltage ($V_\pi$). To prevent generated EM from leaking outside the device, placing it in an EM-shielding enclosure, rack, or room is effective, as verified in this experiment. To prevent leaked EM from carrying useful information, one can adopt a driving scheme with high modulation signal symmetry with respect to the states or employ jamming by superimposing uncorrelated EM waves. A comprehensive combination of these measures is necessary for countermeasures against side-channel attacks, and establishing standards to verify their effectiveness is desirable. This report quantitatively identifies the possibility of EM wave leakage side channels and presents countermeasures. These results provide effective guidelines for establishing standards to construct trusted nodes in QKDN.

In other words, by implementing appropriate countermeasures against EM wave leakage—while considering the specifications required by QKD system users and a trusted node operator, as well as implementation costs—it is possible to realize a practically secure QKDN.

## ACKNOWLEDGMENT

This work was partly supported by the Ministry of Internal Affairs and Communications (MIC), Japan, under the commissioned R&D project "Research and Development for Early Social Implementation of Quantum Cryptography Communication Networks" and the Strategic Innovation Program "Photonics and Quantum Technology for Society 5.0" of the Council for Science, Technology and Innovation, Cabinet Office, Government of Japan (administered by QST). The author expresses gratitude for assistance in experiments and fruitful discussion with Mr. Yoshino and Ms. Seino of NEC.

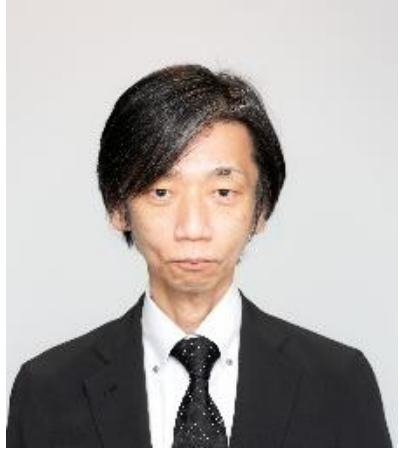

**MIKIO FUJIWARA** received B.S. and M.S. degrees in electrical engineering and a Ph.D. in physics from Nagoya University, Nagoya, Japan, in 1990, 1992, and 2002, respectively. He has been involved in R&D activities at NICT (previously called CRL, Ministry of Posts and Telecommunications of Japan) since 1992.

His main research interests include quantum key distribution (QKD) and QKD network applications.

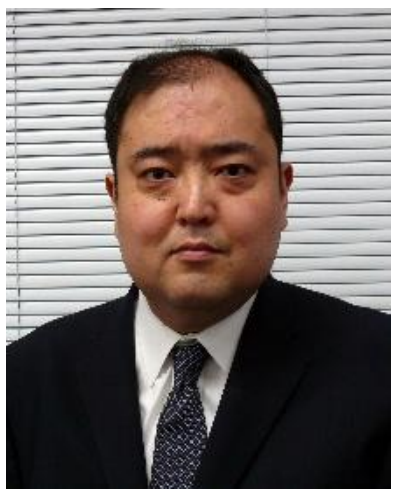

**KATSUMI FUJII** (Member, IEEE) received the B.E., M.E., and Ph.D. degrees in electronic engineering from The University of Electro-Communications, Tokyo, Japan, in 1996, 1998, and 2001, respectively.

From 2001 to 2006, he was a Research Associate with the Research Institute of Electrical Communication, Tohoku University, Sendai, Japan. Since April 2006, he has been with the National Institute of Information and Communications Technology (NICT), Tokyo. He has been engaged in research on EMC measurement technology and involved in the activities of the international special committee on radio interference (CISPR).

Dr. Fujii is a Senior Member of the Institute of Electronics, Information and Communication Engineers (IEICE) of Japan, and a member of the Institute of Electrical Engineers of Japan (IEEJ).

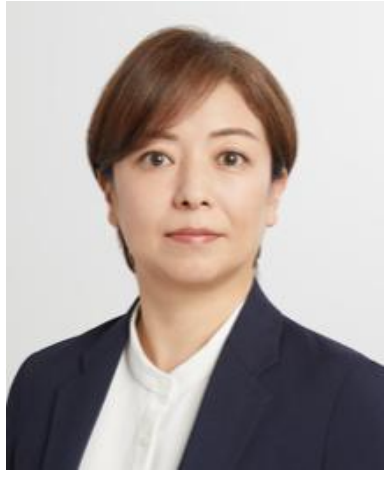

**FUMIE ONO** received the B.E., M.E., and Ph.D. degrees in electrical engineering from Ibaraki University, Mito, Japan, in 1999, 2001, and 2004, respectively. From 2004 to 2006, she was a Research Associate with the Faculty of Engineering, Tokyo University of Science. From 2006 to 2011, she was an Assistant Professor with the Division of Electrical and Computer Engineering, Yokohama National University. She is currently the Initiative Lead of the AI Research and Development Promotion Unit, National Institute of Information and Communications Technology (NICT), Tokyo, Japan. Her current interests include artificial intelligence, AI for Science, foundation models, and AI-enabled communication and network systems. She received the Young Engineer Award from IEICE in 2006 and the YRP Award in 2007.

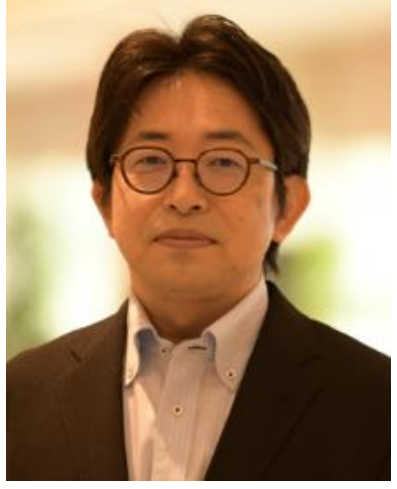

**MASAKAZU ONO** (Member, IEEE) received the B.S., M.S., and Ph.D. degrees in distributed synchronization on mobile networks from Tokyo Denki University, Tokyo, Japan, in 2003, 2005, and 2009, respectively. Since 2008, he has been involved in research, development, and deployment of communication, information, and security systems in the defense and national security domain at NEC Corporation. Since 2023, he has been seconded to the National Institute of Information and Communications Technology (NICT), where his work focuses on advancing research and development toward the societal implementation of quantum cryptography.

Dr. Ono is a member of International Information System Security Certification Consortium (ISC)² and a member of the Institute of Electronics, Information and Communication Engineers (IEICE) of Japan.

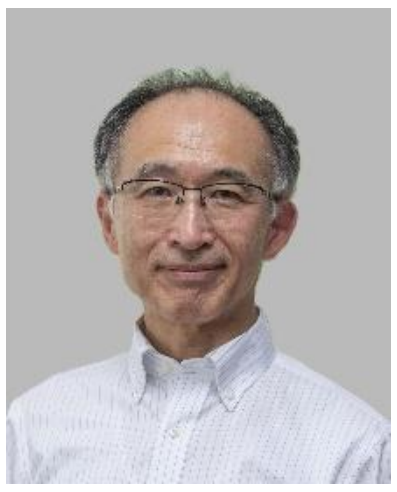

**AKIHISA TOMITA** received a B.S. and an M.S. in physics, and a Ph.D. in electronics, all from the University of Tokyo. From 1984 to 2010, he worked at NEC. He also served as a visiting researcher at AT&T Bell Laboratories from 1991 to 1992 and as the team leader of the JST ERATO Quantum Computation and Information Project from 2000 to 2010. From 2010 to 2025, he was a professor at the Graduate School of Information Science and Technology, Hokkaido University. He then joined NICT in 2025. Since 1998, he has been engaged in research on quantum information technology.

Dr. Tomita is a member of the Physical Society of Japan, the Institute of Electronics, Information and Communication Engineers, and Optica.